# In-fiber measurement of erbium-doped ZBLAN $4I_{13/2}$ state energy transfer parameter

Jillian E. Moffatt,[1,2,*] Georgios Tsiminis,[1,2,3] Elizaveta Klantsataya,[1] Ori Henderson-Sapir,[1] Barnaby W. Smith,[4] Nigel A. Spooner,[1,2,4] and David J. Ottaway[1,2,5]

[1]*Institute for Photonics and Advanced Sensing, School of Physical Sciences, The University of Adelaide, Adelaide, South Australia, Australia, 5005*
[2]*CRC for Optimising Resource Extraction, PO Box 403, Kenmore, Queensland, Australia, 4069*
[3]*ARC Centre of Excellence for Nanoscale BioPhotonics School of Medicine, The University of Adelaide, Adelaide, South Australia, Australia, 5005*
[4]*Defence Science and Technology Group, Edinburgh, Adelaide, South Australia, Australia, 5111*
[5]*ARC Centre of Excellence for Gravitational Wave Discovery, OzGrav, The University of Adelaide, Adelaide, South Australia, Australia, 5005*
*\*jillian.moffatt@adelaide.edu.au*

**Abstract:** Erbium-doped ZBLAN is the most commonly used glass for mid-infrared fiber lasers. In this paper we quantify the energy dynamics of the erbium ions which is important for improving the performance of mid infrared fiber lasers. Previous studies have found a discrepancy between the strength of inter-ion energy transfer measured in bulk Er:ZBLAN and the strength required to explain current fiber laser performance. We have measured the energy transfer of the $4I_{13/2} + 4I_{13/2} \rightarrow 4I_{15/2} + 4I_{9/2}$ energy transfer process directly in a range of fibers for the first time.

1. **Introduction**

Erbium-doped ZBLAN (Er:ZBLAN) glass is used for manufacturing mid-infrared fiber lasers because its mid-infrared transmission is high and its maximum phonon energy is low, which prevents the phonon processes from quenching the excited state populations[1]. ZBLAN is a fluoride glass containing zirconium, barium, lanthanum, aluminum, and sodium. Rare earth ion (RE) dopants are easily substituted for zirconium and lanthanum because they have similar atomic radii. This means the glass can be doped with RE ions at high percentages—up to 4 mol% without glass deformation effects[1] and up to 7 mol% while still maintaining the transmission and phonon energy properties. While in general ZBLAN is an ideal medium for fiber lasers, it is known to allow similar ions to form clusters in the glass structure[2]. In addition, ZBLAN glass will undergo crystallization if the conditions of its formation are not tightly controlled. A study in 2000 found that more than 50% of erbium ions in Er:ZBLAN fibers are in clusters when the fiber is doped at greater than 1 mol%[2]. While ZBLAN production procedures have improved since this time, ZBLAN crystallization is still an issue being studied today[3].

Quantifying the energy dynamics (absorption properties, fundamental lifetimes, energy transfer properties) of the erbium ions will provide new insight into the performance of Er:ZBLAN based fiber laser leading to enhanced performance [4]. One notable property is the inter-ion energy transfer parameter. Ions in an excited state can exchange energy with nearby ions via resonant interactions[5]. Energy transfer parameters are dependent on the host lattice because two RE ions need to be in close proximity to undergo energy transfer, meaning these parameters are dependent on dopant percentage and clustering rates. This makes energy transfer parameters hard to calculate theoretically for complex glass environments such as ZBLAN. Many energy levels in erbium doped ZBLAN have millisecond lifetimes[6, 7]. These comparatively long lifetimes increase the probability of energy transfer even if the process requires both ions to be in excited states.

Numerical models aid laser design and assess variables such as laser excitation power, fiber dopant levels, and core diameter. These models require accurate process parameters, usually gained from bulk glass measurements. One of the strongest energy transfer process in Er:ZBLAN, and therefore one of the most well-studied, is the $^4I_{13/2} + ^4I_{13/2} \rightarrow ^4I_{15/2} + ^4I_{9/2}$ ($W_{11}$) transition. This process takes two ions in the first excited state and transfers one to a higher state ($^4I_{9/2}$) and drops the other to the ground state ($^4I_{15/2}$). This transition is problematic for erbium lasers operating around 1.5 μm and is advantageous for those operating around 2.7 μm. A discrepancy currently exists between the strength of the inter-ion energy transfer processes required to theoretically explain fiber laser output powers[8] and the strength directly measured in bulk ZBLAN[4].

In 2011 Gorjan et al.[8] modelled the decay and energy transfer processes, and fitted their results to output and emission data from a 2.7 μm 6 mol% fiber laser. They varied energy transfer parameters from zero to the highest bulk glass parameters measured. Three laser regimes were modelled: "zero" at near zero energy transfer, "weakly interacting" at $W_{11}$ parameters less than $0.5 \times 10^{-17}$ $cm^3s^{-1}$, and "strongly interacting" above this level.

The 2.7 μm fiber laser Gorjan et al. used followed the behavior of the "weakly interacting" energy transfer regime, whereas bulk glass measurements implied energy transfer parameters in the strongly interacting regime[8]. They theorized that erbium ions cluster in bulk glass, thereby increasing the energy transfer. There are, however, reports of significant clustering in high-doped Er:ZBLAN fibers[2]. The energy transfer regime will likely depend on fiber dopant density[2] and individual fiber construction and history[8].

After Gorjan's paper, Jackson et al. tested the weakly and strongly interacting regimes on multiple published laser output specifications, and found that modelling these outputs agreed with the weakly interacting regime with fibers from 1 to 8 mol%[9]. Since the mid 2010s, generally a pragmatic, qualitative approach to energy transfer has been applied, with low doping regimes ($< 1$ mol%) used to suppress energy transfer[10-12], and high doping regimes ($> 6$ mol%) used to enhance it[13, 14]. Numerical models choose $W_{11}$ numbers in the weakly interacting regime[7, 15, 16], but likely due to the high fiber-to-fiber variability of parameters[12] measuring specific energy transfer parameters has not been attempted. Energy transfer still appears as a problem during laser development[17, 18], however, and there remains a need for more accurate parameters to aid new laser design.

The strength of the $W_{11}$ parameter has previously been calculated from laser behavior and the proportion of fluorescence emission from different erbium states[8]. It has not previously been calculated in fiber from direct measurements of the states involved in the $W_{11}$ process. We describe here an experiment to measure energy transfer parameters directly in fiber, and provide fits of the $W_{11}$ parameter for fibers of different doping densities and differing manufacturers.

2. **Theoretical and experimental parameters**
   *3.1 Rate equation model*
   Our experiment analyses the time-resolved luminescence from the erbium $4I_{13/2}$ and $4I_{11/2}$ states when exciting the $4I_{13/2}$ state. We excite the $4I_{13/2}$ state far from the excited state absorption (ESA) peak, which suppresses ESA and leaves energy transfer as the only upconversion process available. This is checked via observation of emission at 980 nm, which has zero detected counts at time $= 0$. The width of the excitation pulse is much smaller than the lifetime of the states (5 ns opposed to ms), so photon absorption can be neglected in the model. Other assumptions of the model are:

- Branching ratios from Bogdanov[19] show that 99 % of the ions in the $4I_{9/2}$ state decay to the $4I_{11/2}$ state with a lifetime of around 8 μs[6]. As this lifetime is much smaller than the lifetime of the $4I_{11/2}$ state, these two states in the model are merged. This assumption is assumed valid as a more complex model dividing these two states did not change the fit to the $W_{11}$ parameter.
- There was no detectable three-photon upconversion or three-photon absorption occurring; a conclusion drawn as no visible emission was detected during experiments.
- The energy transfer process $4I_{11/2} + 4I_{11/2} \rightarrow 4F_{7/2} + 4I_{15/2}$ ($W_{22}$) does not occur significantly during the experiment. This is considered valid due to the lack of detected visible emission. This would also suppress other energy transfer processes such as $4H_{11/2} + 4I_{15/2} \rightarrow 4I_{9/2} + 4I_{13/2}$ ($W_{50}$) and $4F_{9/2} + 4F_{11/2} \rightarrow 4H_{11/2} + 4I_{13/2}$ ($W_{42}$).

With the assumptions above, we use a simplified rate equation to model the behavior of the Er:ZBLAN fiber:

$$\frac{dN_o}{dt} = N_1{}^2 W_{11} + N_1 A_{10} + N_2 A_{20}$$
$$\frac{dN_1}{dt} = -2N_1{}^2 W_{11} - N_1 A_{10} + N_2 A_{21} \quad (1)$$
$$\frac{dN_2}{dt} = N_1{}^2 W_{11} - N_2 A_{21} - N_2 A_{20}$$

Where $N_0$ is the population density of the $4I_{15/2}$ state, $N_1$ is the population density of the $4I_{13/2}$ state, $N_2$ is the population density of the combined $4I_{9/2}$ and $4I_{11/2}$ states, $W_{11}$ is the energy transfer parameter, $A_{10}$ is the fundamental lifetime of the $4I_{13/2}$ state, and $A_2$ is the fundamental lifetime of the $4I_{11/2}$ state from which $A_{2x}$ values are derived.

*2.2 Experimental setup*

Experiments were conducted using an Edinburgh Instruments F980 Spectrofluorimeter, with an Opotek "Opolette" 20 Hz 5 ns OPO excitation source and a Hamamatsu liquid-nitrogen cooled R5509-72 photomultiplier for detection. The OPO was tuned to 1471 nm for excitation, in order to avoid excited-state absorption from $4I_{13/2} \rightarrow 4I_{9/2}$, which shows as a double peak centered near 1660 nm (FWHM ~ 50 nm) in phosphate glass[20]. A short length of erbium-doped ZBLAN fiber (typically 3-5 cm in length) was placed on an XYZ stage inside the measurement chamber. The outer polymer cladding and jacketing was removed from the tip of the fiber which was aligned end-on to the focused beam of the OPO (see Fig. 1.). The pulse energy was varied by placing a series of neutral density filters into the path of the beam prior to focusing.

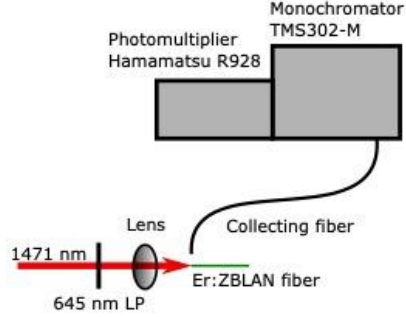

*Fig. 1. Experimental setup*

The fluorescence from the tip of the doped fiber was collected by a silica fiber with a 1 mm diameter core. The light collected by this fiber was incident on the slit of the spectrofluorimeter monochromator. The monochromator was set to pass a center wavelength of 1530 nm for detection of $4I_{13/2}$ → $4I_{15/2}$ emission, and a center wavelength of 980 nm for $4I_{9/2}$ → $4I_{15/2}$ emission. Time dependent measurements over 40 ms were made for successive laser pulses and integrated until the signal was visible above the noise, taking between 15 minutes (1.8x10$^4$ pulses) for highly doped fibers at 1530 nm and 27 hours (approximately 1.9x10$^6$ pulses) for low doped fibers at 980 nm.

Measurements were made using commercial erbium doped fibers with varying core and inner cladding diameters. The erbium content ranged from 1 to 7 mol%. Two 1 mol% and two 7 mol% fibers from different manufacturing batches were used in order to assess repeatability of experiments and fiber-to-fiber variability.

*2.3 Initial population*
The initial value of the population $N_1$ is defined as the proportion of ions excited to the $4I_{13/2}$ state via the 5 ns laser pulse. When far from saturation, this can be calculated using the following equation:

$$N_{1(initial)} = N_g \frac{E}{hv} \frac{\sigma_{1471}}{a}$$

(2)

Where $N_g$ is the ion density of the fiber core, E is the pulse energy incident on the fiber core, *h* is Plank's constant, *a* is the area of the fiber core, $\sigma_{1471}$ is the absorption cross-section at the excitation wavelength, and *v* is the excitation frequency.

*2.4 Energy per pulse*

The laser was not tightly focused into the fiber core, so that beam intensity variations across the core would be minimized. A knife-edge measurement, which takes energy measurements as the beam is slowly occluded in the x direction, was performed at the yz position at which the fiber was placed to calculate the alignment and variation of the beam. Fitting this to a Gaussian beam profile allowed a calculation of the energy incident on the fiber core.

*2.5 Absorption cross-section*

The absorption cross-section was taken from bulk glass measurements published previously[21]. Individual absorption cross-section measurements were also calculated for each fiber from fluorescence spectra via the McCumber relationship[22] and the Füchtbauer-Ladenburg equation[23]. Absorption cross-section measurements calculated in this way are more inaccurate, but provided a fair indication of fiber-to-fiber variability and were used in error analysis. Most fibers had similar absorption cross-sections, with some small exceptions shown in Fig. 2.

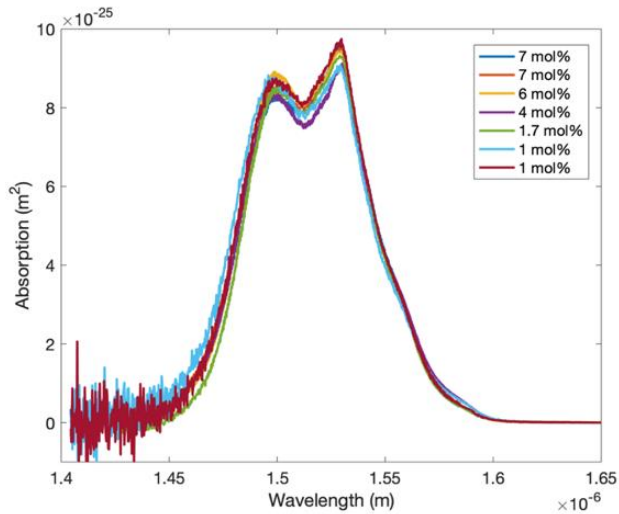

*Fig. 2. Absorption cross-section data for all fibers gained from emission spectra via the McCumber relationship and the Füchtbauer-Ladenburgh equation.*

*2.6 Lifetime (1/A)*

The reciprocal lifetimes $A_1$ and $A_2$ were found experimentally by fitting to the tail of time resolved emission at 1530 nm excited by 1471 nm and at 980 nm excited by 990 nm. Lifetimes differed only slightly between fibers and the results agreed well with literature[6]. The branching ratio for $A_{21}$ and $A_{20}$ was taken from Malouf et al.[7].

*2.7 Analysis*

Time resolved data at 1530 nm was first normalized to the calculated $N_{1(initial)}$ value, then fitted to a numerical solution of Eq. 1 using the Matlab® differential equation solver. The energy transfer parameter $W_{11}$ was the only unknown variable. Time resolved data at 980 nm were normalized with a factor calculated by matching 7 mol% 980 nm results to 7 mol% 1530 nm results (Fig. 3.d). Fitting errors were small, and more accurate estimates of errors were found by taking into account variations in fundamental lifetime and absorption cross-sections.

### 3. Results and discussion

**Table 1: $W_{11}$ values gained from fits to time-resolved measurements**

| Erbium Percentage (mol%) | Core diameter (μm) | $W_{11a}$ using 1530 nm $10^{-18}cm^3s^{-1}$ | $W_{11a}$ using 980 nm[b] $10^{-18}cm^3s^{-1}$ |
|---|---|---|---|
| 1 | 16.5 | 0.2 (0, 1.9) | 1.3 (1.0, 2.1) |
| 1[c] | 16.5 | 0.3 (0, 2.0) | - |
| 1.7 | 10 | 0.3 (0.1, 0.8) | 0.9 (0.6, 1.4) |
| 4 | 14 | 2.5 (1.7, 4.4) | - |
| 6 | 25 | 4.2 (2.5, 6.4) | - |
| **Weakly interacting regime < 5 × $10^{-18}cm^3s^{-1}$ < Strongly interacting regime (Gorjan et al., 2011)** | | | |
| 7 | 16.5 | 11 (7.5, 18) | - |
| 7[c] | 16.5 | 15 (9.1, 29) | Used for calibration |

*[a]Error ranges are variations of fit results when changing lifetime (s) and absorption cross-section (m2) values by errors (1 σ).*
*[b]980 nm emission was only calculated for 1 mol% and 1.7 mol% fibers.*
*[c]From the same manufacturer but different reels.*

The measured $W_{11}$ values are shown in Table 1. Both 7 mol% fibers are in the strongly interacting regime, and the lower-mol% fibers are in the weakly interacting regime (the 6 mol% fiber could be in either regime within errors). Fiber-to-fiber variability in fibers of the same mol% was within measurement errors.

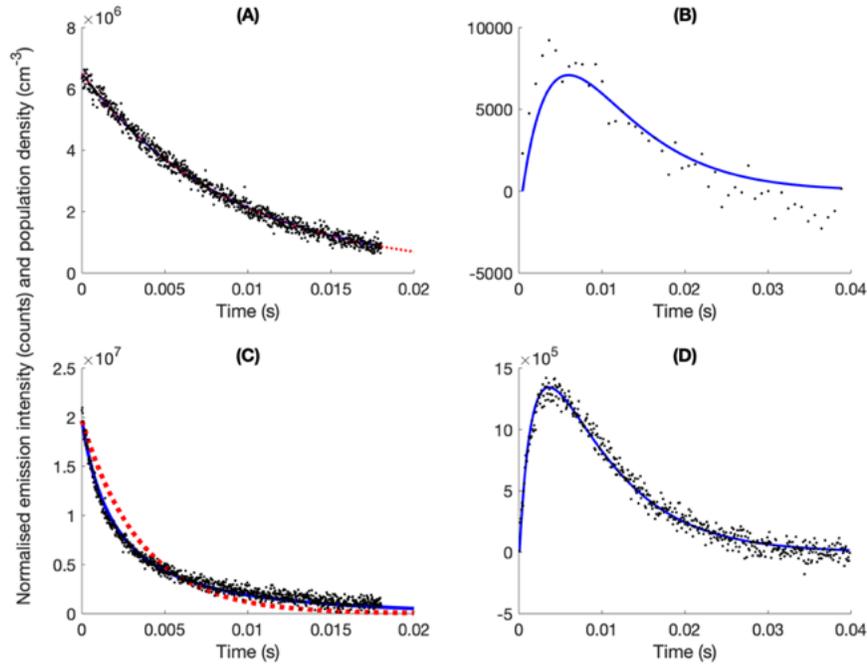

*Fig. 3. Time-resolved data and fits to data for a) 1 mol% fiber at 1530 nm; b) 1 mol% fiber at 980 nm; c) 7 mol% fiber at 1530 nm; and d) 7 mol% fiber at 980 nm (used for calibration). Blue solid line depicts fit to model using Eq. 1.; red dashed line depicts fit to single exponential.*

Fitting the 1530 nm emission from 1 mol% and 1.7 mol% fibers resulted in large errors as the deviation from a simple exponential was very low (see Fig. 3.a), making the fit unstable. As a check, fits were attempted on the 980 nm emission, which is more reliant on the $W_{11}$ parameter. Due to the very low counts at 980 nm these fits also had large errors, but confirmed the range of $W_{11}$ parameters for very low-doped fiber. While it is expected that, with all other things being equal, the higher-doped fiber would have a stronger $W_{11}$ component, this does not appear to be the case for the 1.7 mol% fiber. 1 mol% to 1.7 mol% fiber variability, however, was already seen in the calculated absorption cross-sections in Fig. 2, and so a difference in the fiber-to-fiber $W_{11}$ value relationship is not unexpected.

## 4. Conclusions

$W_{11}$ parameters were measured directly in Er:ZBLAN fiber for the first time from time resolved data. The energy transfer parameter moved from the weakly interacting regime at 1 mol% to the strongly interacting regime at 7 mol%. While there was a reasonable amount of fiber-to-fiber variability, fibers from the same manufacturer are not expected to change regimes unless they are doped near 6 mol%. These findings have implications for the modelling and design of erbium-doped ZBLAN fiber lasers.

## Funding

The authors disclosed receipt of the following financial support for the research, authorship, and/or publication of this article: CRC for Optimising Resource Extraction grant (project number P1-005); an Australian Government Research Training Program (RTP) Scholarship; and a LIEF grant (code LE140100042).

## Acknowledgements

The authors wish to thank Deeksha Beniwal and Prof. S D Jackson for supplying fiber samples.

## Disclosures

The authors declare no conflict of interest.